\documentclass[conference]{IEEEtran}
\IEEEoverridecommandlockouts
\usepackage{cite}
\usepackage{amsmath,amssymb,amsfonts}
\usepackage{algorithmic}
\usepackage{graphicx}
\usepackage{textcomp}
\usepackage{xcolor}
\def\BibTeX{{\rm B\kern-.05em{\sc i\kern-.025em b}\kern-.08em
    T\kern-.1667em\lower.7ex\hbox{E}\kern-.125emX}}
\DeclareRobustCommand*{\IEEEauthorrefmark}[1]{%
  \raisebox{0pt}[0pt][0pt]{\textsuperscript{\footnotesize\ensuremath{#1}}}}
\begin{document}

\title{Gradient of White Matter Functional Variability via fALFF Differential Identifiability}
\author{
\IEEEauthorblockN{Xinle Chang\IEEEauthorrefmark{1},
Yang Yang\IEEEauthorrefmark{1},
Yueran Li\IEEEauthorrefmark{1}, 
Zhengcen Li\IEEEauthorrefmark{1},
Haijin Zeng\IEEEauthorrefmark{2}, and
Jingyong Su\IEEEauthorrefmark{1*}}
\IEEEauthorblockA{\IEEEauthorrefmark{1}School of Computer Science and Technology, Harbin Institute of Technology, Shenzhen, China}
\IEEEauthorblockA{\IEEEauthorrefmark{2}Department of Psychiatry, Harvard University, Cambridge, USA}
\thanks{This paper was accepted by IEEE International Conference on Bioinformatics and Biomedicine (BIBM) 2025.}
}


\maketitle

\begin{abstract}
Functional variability in both gray matter (GM) and white matter (WM) is closely associated with human brain cognitive and developmental processes, and is commonly assessed using functional connectivity (FC). 
However, as a correlation-based approach, FC captures the co-fluctuation between brain regions rather than the intensity of neural activity in each region. 
Consequently, FC provides only a partial view of functional variability, and this limitation is particularly pronounced in WM, where functional signals are weaker and more susceptible to noise.
To tackle this limitation, we introduce fractional amplitude of low-frequency fluctuation (fALFF) to measure the intensity of spontaneous neural activity and analyze functional variability in WM. 
Specifically, we propose a novel method to quantify WM functional variability by estimating the differential identifiability of fALFF. 
Higher differential identifiability is observed in WM fALFF compared to FC, which indicates that fALFF is more sensitive to WM functional variability. 
Through fALFF differential identifiability, we evaluate the functional variabilities of both WM and GM, and find the overall functional variability pattern is similar although WM shows slightly lower variability than GM.
The regional functional variabilities of WM are associated with structural connectivity, where commissural fiber regions generally exhibit higher variability than projection fiber regions. 
Furthermore, we discover that WM functional variability demonstrates a spatial gradient ascending from the brainstem to the cortex by hypothesis testing, which aligns well with 
the evolutionary expansion. 
The gradient of functional variability in WM provides novel insights for understanding WM function. 
To the best of our knowledge, this is the first attempt to investigate WM functional variability via fALFF. 
Our code is available at https://github.com/Xinle-Chang/WM-fALFF-Idiff-Gradient.
\end{abstract}

\begin{IEEEkeywords}
Functional variability, White matter fMRI, fALFF, Differential identifiability, Gradient
\end{IEEEkeywords}

\section{Introduction}
\label{sec:intro}

Functional magnetic resonance imaging (fMRI) based on the blood oxygenation level dependent (BOLD) signal has been a predominant technique for detecting neural activity in the human brain~\cite{biswal2012resting,ding2013spatio,fMRI_logothetis2008we}. Gray matter (GM), an area with dense neurons and synapses, has been extensively studied for its cognitive functions. In comparison, white matter (WM) consists primarily of nerve fibers responsible for signal transmission and exhibits lower cerebral blood flow along with distinct neural activity patterns~\cite{wm_review2,wm_review1}. The significance of WM BOLD signals remained unclear for an extended period, with early studies often treating them as noise or confounding factors~\cite{filley2016white,parvizi2009corticocentric}. However, recent studies have provided compelling evidence of an intrinsic functional organization within WM, similar to GM, suggesting that WM BOLD signals can reflect meaningful neural activity indeed~\cite{ding_pnas,peer2017,zy_tmi}.

Genetic and environmental factors can affect different brain systems to varying degrees, leading to a spectrum of brain function differences, which is known as functional variability~\cite{1998Comparative}.
Functional variability in WM, akin to that observed in GM, is an important marker for validating the effectiveness of functional signals within WM.
Typically, functional variability can be observed and quantified by analyzing the functional connectome~\cite{connectomic_fingerprints} of individual brains. 
Research has revealed that GM functional variability is significantly correlated with evolutionary expansion degree and cortical sulcal depth, with heteromodal cortices exhibiting higher variability compared to unimodal cortices~\cite{Mueller_IVF}. 
Furthermore, WM functional variability and its genetic underpinnings have been identified through functional connectivity (FC) analysis~\cite{IVF_WMFC}. 
However, conventional FC focuses on correlations between brain regions or voxels, overlooking the amplitude and frequency features of the BOLD signals emitted by spontaneous neural activities of these regions or voxels.

To tackle the limitations, the differential identifiability ($I_{diff}$) of fractional amplitude of low-frequency fluctuation (fALFF) is proposed in this study to evaluate WM functional variability. 
fALFF quantifies the intensity of brain activity during resting-state fMRI, defined as the ratio between the power spectrum of low-frequency oscillations and the full frequency range~\cite{fALFF_origin}. Unlike indicators representing co-fluctuation between BOLD signals, such as FC, fALFF focuses on the spontaneous low-frequency fluctuations of BOLD signals derived from neural activities. It partially filters out high-frequency noise, making it more stable and resistant to confounding factors. Consequently, fALFF is widely applied in the diagnosis and research of various brain disorders~\cite{hu2025dynamic,lv2025research,yang2025impact_mdd}. 
Recent studies have highlighted the biophysical significance of low-frequency WM BOLD signal fluctuations encoding the neural activities~\cite{JIGONGJUN,WMFC}, suggesting that fALFF is an effective indicator for detecting neural activities within WM. Functional variability can be quantified by the $I_{diff}$, which is conventionally defined as the difference between the mean intra-subject similarity and the mean inter-subject similarity of the FC matrices~\cite{idiff}. 
By shifting the focus from FC to fALFF similarity, we propose a novel metric fALFF $I_{diff}$ for a more comprehensive analysis of WM functional variability.

\begin{figure*}[htb]
    \centering
    \centerline{\includegraphics[width=0.8\textwidth]{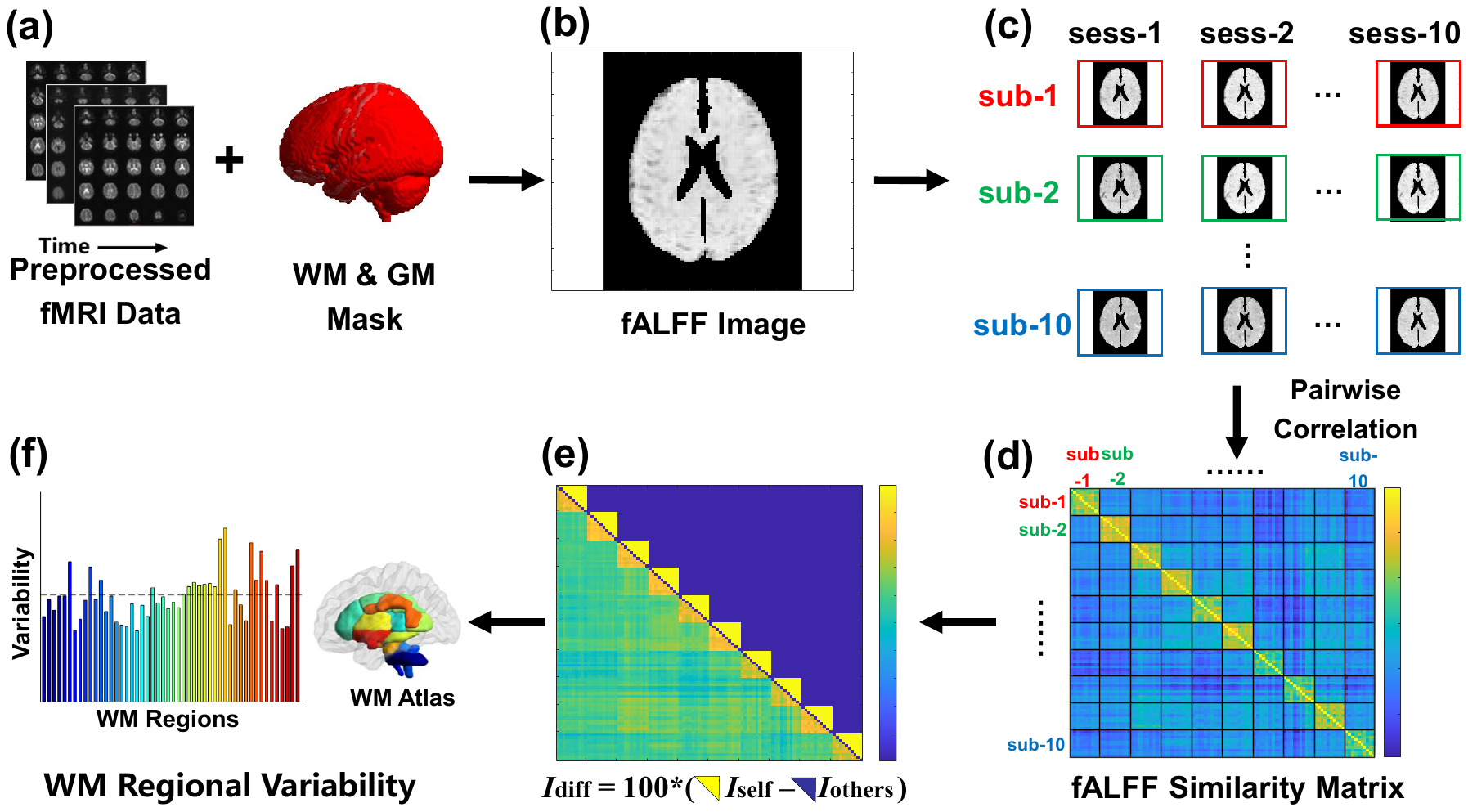}}
    \caption{\textbf{Overview of fALFF $I_{diff}$ analysis framework for assessing functional variability.} (a-b) A group-level mask is applied to select WM and GM voxels, in which BOLD-fMRI signals are used to compute fALFF images. (c-d) A fALFF Similarity Matrix is constructed by pairwise correlations of fALFF images across scans of subjects. (e)$I_{diff}$ is computed by subtracting the average inter-subject similarity from the average intra-subject similarity. (f) Regional functional variability can be quantified and analyzed by substituting the group-level mask with each region in the ICBM-DTI-81 atlas.}
    \label{fig:schematic}
\end{figure*}


In summary, the main contributions of this work are as follows: 
\begin{enumerate}
\item[$\bullet$] We propose a novel method for evaluating functional variability at various scales by creatively linking fALFF with $I_{diff}$. 
\item[$\bullet$] By evaluating the global functional variabilities of both WM and GM, we find that although WM exhibits slightly lower functional variability than GM, the overall variability pattern is similar.
\item[$\bullet$] Our investigation of the relationship between WM functional variability and structural connectivity reveals that commissural fiber regions tend to show higher functional variability than projection fiber regions.
\item[$\bullet$] We discover that WM functional variability follows a spatial gradient, increasing from the brainstem to the cortex, which is consistent with brain evolutionary expansion. The existence of this spatial gradient is statistically validated using hypothesis testing procedures.
\end{enumerate}

\section{Materials and Methods}
\label{sec:method}

\subsection{Data Acquisition and Preprocessing}
\label{sec:datamethod}

In this study, we utilized data from two independently acquired publicly available datasets: Human Connectome Project (HCP)~\cite{hcpdataset} and Midnight Scan Club (MSC) dataset~\cite{MSC}.
These datasets encompass three common types of imaging sessions: resting-state fMRI (rsfMRI), T1-weighted image (T1w), and diffusion image (dMRI). 
Relative to HCP, MSC contains a smaller number of participants but offers richer intra-subject data through repeated scanning sessions.

The MSC dataset focuses on detailed feature descriptions of 10 participants (50\% female, mean age 29.1 ± 3.3 years). During the MSC data collection period, each participant underwent 12 independent scanning sessions, including 2 scans to obtain T1w images, T2w images, and other structural data, as well as 10 scans for collecting resting-state and task-based fMRI data. The 10 rsfMRI scans for each participant utilized a Gradient-echo EPI sequence (scan duration = 30 minutes, repetition time (TR) = 2200 ms, echo time (TE) = 27 ms, flip angle = 90°), during which participants were instructed to fixate on a plus sign and remain awake. All images in this dataset were acquired using a 3T Siemens Trio scanner.

The HCP dataset comprises rsfMRI data from 100 unrelated adults(54\% female, mean age = 29.1 ± 3.67 years, age range: 22-36 years). 
To ensure robust statistical inference, the HCP intentionally selected unrelated individuals, thereby minimizing genetic homogeneity and its potential confounding effects.
The rsfMRI data were acquired using a Gradient-echo EPI sequence on a 3T Siemens Connectome Skyra using a 32-channel head coil (scan duration = 14:33 minutes, TR = 720 ms, TE = 33.1 ms, flip angle = 52 degrees, isotropic voxel resolution = 2 mm, multi-band factor = 8)~\cite{hcp_resting}, with eyes of participants open, fixating on a cross during the scan. 
Each participant underwent four 15-minute rsfMRI scans (left-to-right and right-to-left phase coding in sessions REST1 and REST2) in two days.


Data quality is influenced by numerous factors, including the scanning system, acquisition standards, and the characteristics of the test population~\cite{esteban2019fmriprep}.
Therefore, dataset-specific preprocessing pipelines were implemented.
In the HCP dataset, the rsfMRI data we utilized had been processed through a dataset-specific preprocessing pipeline and accurately coregistered to the MNI152 standard space, ensuring high-quality alignment for subsequent analysis.
Compared to the HCP dataset, the raw rsfMRI data from MSC dataset exhibited comparatively lower image quality and was therefore preprocessed using the statistical parametric mapping software toolkit SPM12~\cite{bazay2022preprocessing} to enhance its suitability for analysis.The following steps were adopted:
(\MakeUppercase{\romannumeral 1})\textbf{ Slice timing: }Given the specified slice order, TR, and other pertinent parameters, we applied slice timing correction to a series of 100 resting-state fMRI scans to alleviate the impact of sampling time variations among different slices;
(\MakeUppercase{\romannumeral 2})\textbf{ Motion correction: }Subsequently, realignment with default settings is conducted on all functional images to correct for minor head movements; 
(\MakeUppercase{\romannumeral 3})\textbf{ Spatial coregistration and normalization: }They are crucial to minimize variations in brain shape and structure among individuals. Functional images were coregistered to corresponding T1W structural images, which in turn are registered to a standard template space. Then the transformation matrix obtained was applied to all functional images, resulting in preprocessed fMRI data that conformed to the MNI152 standard space;
(\MakeUppercase{\romannumeral 4})\textbf{ Spatial smoothing: }Spatial smoothing was achieved by applying an empirically sized Gaussian kernel. 
Bandpass filtering is typically applied during fMRI preprocessing to reduce noise and retain meaningful signals. However, because fALFF quantifies the ratio of low-frequency oscillations to the entire frequency spectrum, bandpass filtering would compromise its validity. Consequently, no bandpass filtering was applied in this study.

Due to the close anatomical integration of WM and GM, distinguishing between WM and GM voxels remains an important challenge.
The SPM12 provides a tissue probability maps (TPM) based on the average brain template in MNI152 standard space, indicating the likelihood that each voxel belongs to specific brain tissue types, such as WM, GM and cerebrospinal fluid (CSF). 
Multiple thresholds were applied to generate suitable masks containing WM or GM voxels.
We observed that voxel count in masks plateaus when the probability threshold drops below 0.8, indicating convergence of the masks.
Therefore, WM and GM masks were generated by applying a threshold of 0.8 to their respective TPMs, effectively excluding voxels with lower classification confidence.


\subsection{Measurement of fALFF Similarity}
\label{sec:falffmethod}

\begin{figure}
    \centering
    \includegraphics[width=0.9\linewidth]{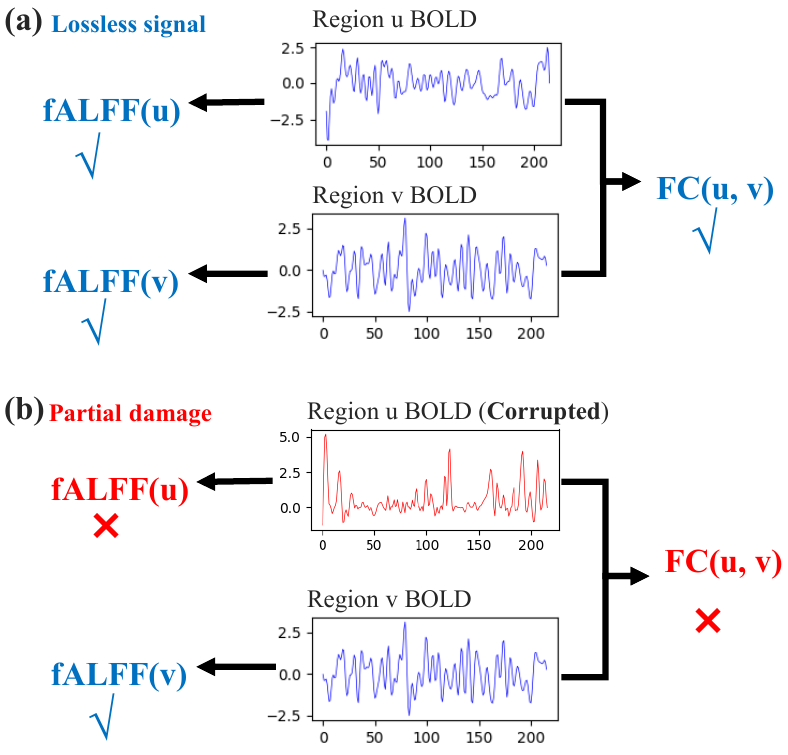}
    \caption{\textbf{Motivation of adopting fALFF rather than FC.} (a) Consider BOLD signals from two regions u and v, as an example. When signals are fully preserved, both FC and fALFF can effectively characterize neural activity. (b) However, if the signals are partially degraded, FC dependending on inter-regional synchrony fails to capture true activity, whereas fALFF, focused on local low-frequency power, continues to reflect the underlying uncorrupted neural signals.}
    \label{fig:fALFF_over_FC}
\end{figure}

Unlike the co-fluctuation pattern revealed by conventional FC~\cite{hutchison2013dynamic}, fALFF predominantly reflects the spontaneous neural activity of local brain regions~\cite{fALFF_origin}. 
FC examines the interrelation of signals across various brain regions, whereas fALFF assesses the intensity of low-frequency signal amplitudes within localized brain regions.
As illustrated in Fig.~\ref{fig:fALFF_over_FC}, using BOLD signals from two representative regions u and v, we demonstrate that while both FC and fALFF perform reliably under ideal conditions, FC becomes invalid when the signals are partially corrupted. In contrast, fALFF remains robust, as it quantifies local signal amplitude independently of inter-regional correlation.
Specifically, fALFF quantifies the proportion of the low-frequency signal components relative to the overall signal in a single voxel, and it can be calculated as:
\begin{equation}
    fALFF_v = \frac{\int_{f_1}^{f_2} \left| F(f) \right| \, df}{ \int_{f_{min}}^{f_{max}} \left| F(f) \right| \, df}
\end{equation}
where $F(f)$ is the signal amplitude at frequency f. [$f_{min}$, $f_{max}$] is the frequency range of the full band, and [$f_1$, $f_2$] is the low-frequency range. We selected the conventional low-frequency range set~\cite{lowfreq_range} $f_1$=0.01Hz, $f_1$=0.08Hz in this study.

Since the similarity between scalar $fALFF_v$ values cannot be directly calculated, as illustrated in Fig.~\ref{fig:schematic} (a-d), we computed the $fALFF_v$ of each voxel individually and combined them into an fALFF image for similarity analysis. 
This fALFF image based on preprocessed fMRI data and a group-level mask is vectored as $\mathrm{fALFF} (s, t)$, where $s (s = 1,2,..., N_s)$ denotes the number of subjects, and $t (t \in 1,2,..., N_t)$ denotes the number of sessions. The $N_s$ and $N_t$ denote total number of subjects and sessions respectively.
For instance,we computed 100 fALFF images corresponding to MSC data from 10 subjects in 10 sessions, so $N_s=10$ and $N_t=10$ in this study.
To verify the functional variability in fALFF images, we generate an fALFF similarity matrix S by computing the pairwise correlation of fALFF images $\mathrm{fALFF} (s, t)$:
\begin{equation}
    S_{ij}=corr(\mathrm{fALFF}(s_p,t_m),\mathrm{fALFF}(s_q,t_n))
\end{equation}
where $corr(.,.)$ denotes the Pearson correlation coefficient of 2 vectors. The variables $p$ and $q$ denote subjects ($p, q=1,2,...,N_s$), while $m$ and $n$ denote sessions ($m, n=1,2,...,N_t$). The $i$ and $j$ denote the elemental corrdinates of S, where $i=(p-1)*10+m$ and $j=(q-1)*10+n$. As illustrated in Fig.~\ref{fig:schematic} (d), the fALFF similarity is symmetric, with its rows and columns primarily organized according to the order of subjects.


\subsection{Quantization of Functional Variability}
\label{sec:idiffmethod}
The diagonal and off-diagonal blocks of fALFF similarity matrix can comprehensively represent the similarity of intra-subject and inter-subject fALFF images respectively~\cite{subject_efc}.
However, the fALFF similarity matrix can not be directly utilized for quantifying WM functional variability, which calls for an effective metric to measure such variability. 
By analyzing the element distribution of fALFF similarity matrix, we find that intra-subject similarities were significantly greater than inter-subject similarities. Thus we introduced $I_{diff}$ to measure WM functional variability by this similarity difference.


The $I_{diff}$ is developed to evaluate functional variability through FC and to select the optimal components in principal component analysis (PCA) reconstruction initially~\cite{idiff}. 
It is inherently well-suited for the analysis of similarity matrices. Consequently, we used $I_{diff}$ to quantify the disparity in similarities between fALFF images within the same subject versus those across different subjects. 
In fALFF similarity matrix, $I_{diff}$ is computed as the difference between the mean intra-subject similarity, denoted as self-identifiability $I_{self}$, and the mean inter-subject similarity, known as others-identifiability $I_{others}$:
{\small
\begin{equation}
I_{self}=E(corr(\mathrm{fALFF}(s_p,t_m),\mathrm{fALFF}(s_p,t_n))), m\neq n
\end{equation}
\begin{equation}
    I_{others}=E(corr(\mathrm{fALFF}(s_p,t_m),\mathrm{fALFF}(s_q,t_n))),p\neq q
\end{equation}
}\begin{equation}
    I_{diff}=I_{self}-I_{others}
\end{equation}
where E(.) denotes the mean function. From a computational perspective, $I_{diff}$ quantifies the difference between the mean similarity values within the diagonal blocks (representing intra-subject similarity) and those in the off-diagonal blocks (representing inter-subject similarity) of the fALFF similarity matrix.

By replacing the group-level mask in Fig.~\ref{fig:schematic} (a) with WM regions in ICBM-DTI-81 atlas~\cite{atals1,atlas2}, we can obtain fALFF $I_{diff}$ values of all WM regions for analyzing the gradient distribution of functional variability. So we initially generated WM regional fALFF $I_{diff}$ values as depicted in Fig.~\ref{fig:schematic} (f). 

\subsection{Hypothesis and Validation of Gradient}
\begin{figure}
    \centering
    \includegraphics[width=0.9\linewidth]{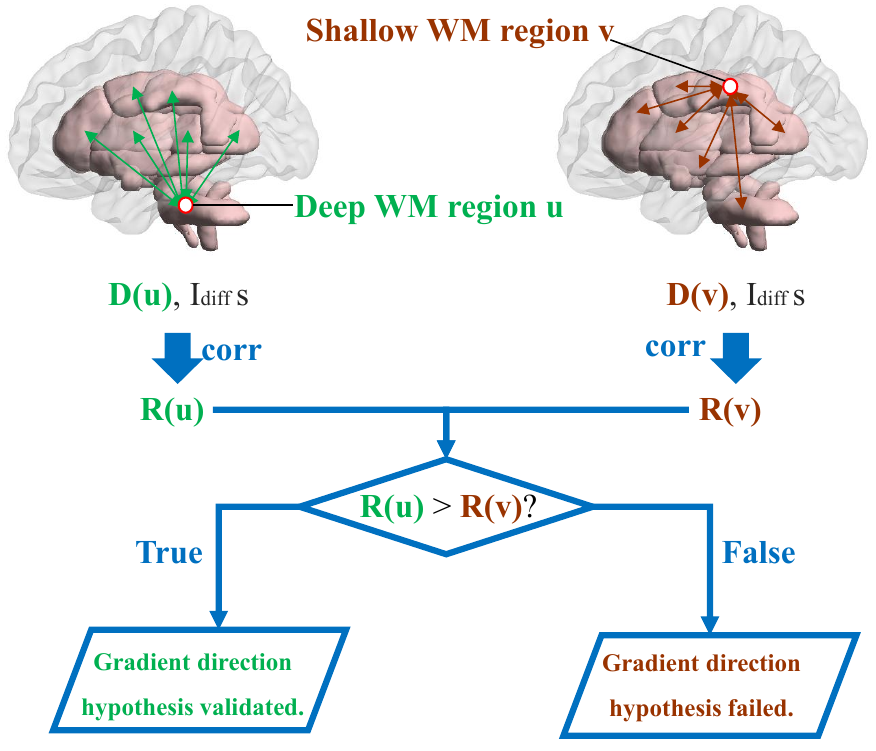}
    \caption{\textbf{Flowchart of hypothesis testing on gradient.} To validate the presence of a spatial gradient in WM functional variability, we compute the relative distance vectors $D(u)$ and $D(v)$ using a deep WM region u and a shallow WM region v as reference regions. We then calculate the correlation coefficients $R(u)$ and $R(v)$ between these distance vectors and the $I_{diff}$ values across WM regions. If $R(u) > R(v)$, it confirms a gradient of increasing functional variability from deep to shallow WM.}
    \label{fig: gradient_testing}
\end{figure}

According to the evolutionary expansion pattern of the human brain, we hypothesized that the functional variability of WM regions is proportional to their anatomical distance from the deepest WM regions—most notably, the brainstem—reflecting a potential spatial gradient in variability.
Put differently, as the distance from the brainstem increases and regions approach the cortical surface, their functional variability correspondingly increases.
As depicted in Fig.~\ref{fig: gradient_testing}, a novel method specifically designed for the stereotaxic architecture of WM is proposed to validate the existence of this gradient.
Utilizing the spatial center coordinates of all WM regions, with WM region $u$ as the reference region, we can compute the relative distances to $u$ of all WM regions and amalgamate them into a vector denoted as $D(u)$. Pearson or Spearman correlation coefficient $R(u)$ of $D(u)$ and fALFF $I_{diff}$ values can be employed to ascertain the presence of substantial spatial gradients in functional variability:
\begin{equation}
    R(u)=corr(I_{diff}s, D(u))
\end{equation}
where $I_{diff}s$ denotes the array WM regional fALFF $I_{diff}$ values. We computed correlation coefficients for all WM regions and checked the overlap between highly relevant WM regions and the brainstem to validate our hypothesis.
To confirm the existence of a functional variability gradient in WM, it is required that only the brainstem or its neighboring deep WM regions show a correlation coefficient above the standard threshold (e.g., $r \geq 0.3$) along with high statistical significance ($p \leq 0.01$).


\section{Experimental Results}
\label{sec:result}

In this study, we conducted a global comparison of functional variability between WM and GM, then performed a regional analysis across WM structures, and ultimately confirmed the existence of a spatial functional variability gradient in WM via hypothesis testing.
Throughout this section, we used data from two independently acquired datasets: the MSC dataset and the HCP dataset. 
Given that the MSC dataset provides a balanced structure with ten participants each scanned ten times, and the HCP dataset offers broader coverage with 100 participants but only two scans per person, we focused our primary analyses on MSC and used HCP for independent validation. Consequently, we obtained consistent conclusions on HCP and MSC datasets.

\subsection{Global Functional Variability}
\label{fig:global_res}

\begin{figure}[htbp]
    \centering
    \includegraphics[width=0.9\linewidth]{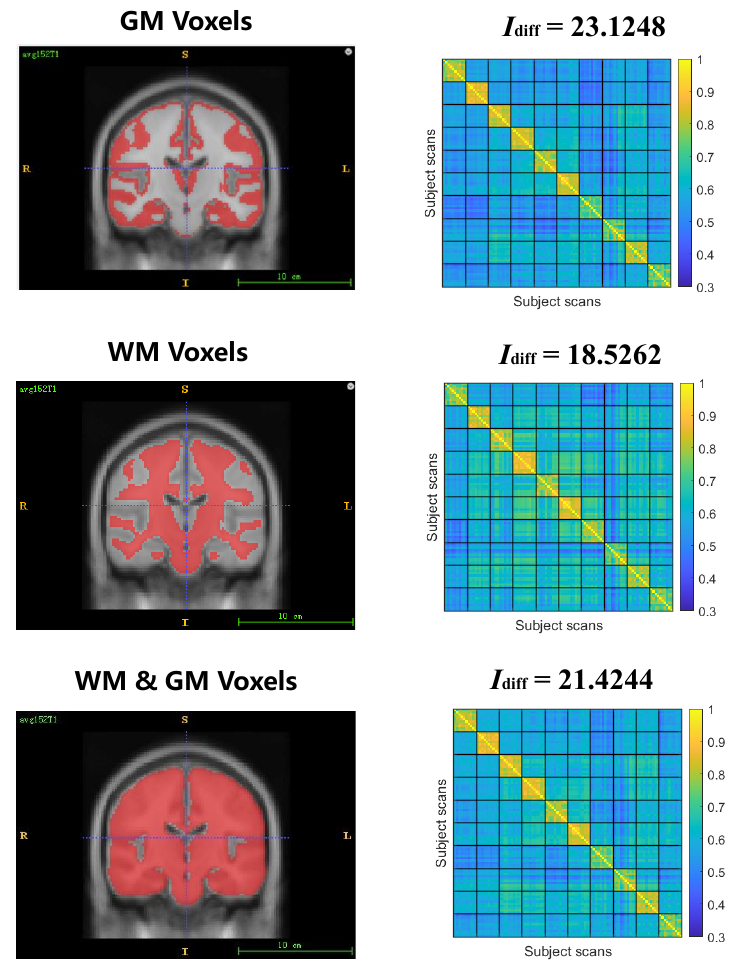}
    \caption{\textbf{Coronal view of WM, GM masks and comparison of GM, WM, WM and GM in fALFF similarity matrix and corresponding $I_{diff}$s.} Longitudinal slices of the 3 masks are visualized, with voxels corresponding to each tissue type highlighted in red. In fALFF similarity, WM has similar yet marginally lower functional variability than GM. The observed increase in $I_{diff}$ from WM to WM-GM mixed regions and then to GM implies a potential functional variability gradient from WM to GM.}
    \label{fig:comparison}
\end{figure}



From the preprocessed fMRI data in the MSC dataset, we initially generated 100 fALFF images corresponding to data from 10 subjects across 10 sessions.
As depicted in Fig.~\ref{fig:comparison}, by utilizing group-level masks for GM, WM, and a combination of both WM and GM, we computed fALFF similarity matrices and corresponding $I_{diff}$ through correlation of the fALFF images. 
Across WM, GM, and WM \& GM, the fALFF similarity matrices consistently show that intra-subject similarity is greater than inter-subject similarity, as reflected by higher values in the diagonal blocks relative to the off-diagonal blocks.
In the comparison of fALFF $I_{diff}$, the highest functional variability is observed in GM, followed by WM \& GM, with WM exhibiting the least variability. 
Consistent with the anatomical and metabolic distinction between GM and WM, we initially observed a monotonic decrease in functional variability from GM to mixed GM/WM voxels and finally to deep WM. 
This gradient likely reflects a combination of higher synaptic activity and vascular density in GM~\cite{parvizi2009corticocentric}, partial-volume contributions at the interface, and the lower-amplitude, delayed hemodynamic response in WM~\cite{wm_review1}.
This finding also aligns with the evolutionary expansion perspective that WM, which emerged earlier, exhibits lower functional variability, whereas GM, a later evolutionary development, shows higher variability~\cite{Mueller_IVF,hill2010similar}.

\begin{figure}[htbp]
    \centering
    \includegraphics[width=0.9\linewidth]{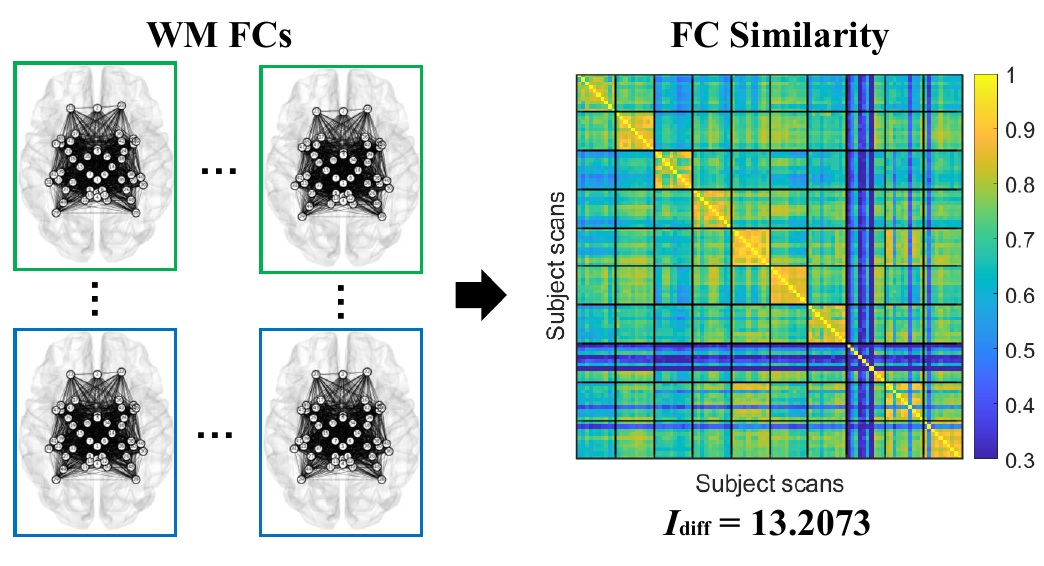}
    \caption{\textbf{WM FCs, corresponding similarity matrix and $I_{diff}$. }WM FCs are derived from the MSC dataset and ICBM-DTI-81 WM atlas. In terms of similarity matrix and $I_{diff}$, FC($I_{diff}$=13.2073) shows lower functional variability than fALFF($I_{diff}$=18.5262). Accordingly, fALFF appears to offer a more robust and sensitive assessment of WM functional variability than FC.}
    \label{fig:wmfc}
\end{figure}

Furthermore, to assess the distinctions between FC and fALFF in measuring 
WM functional variability, based on ICBM-DTI-81 WM atlas and fMRI data, we constructed 100 WM FCs and calculated the corresponding similarity matrix and $I_{diff}$.
As depicted in Fig.~\ref{fig:comparison}, fALFF shows higher $I_{diff}$ than FC in WM, which indicates that fALFF demonstrates superior performance in quantifying WM functional variability. 
Despite the observed overall trend of increasing functional variability from WM to GM, the internal spatial gradient within WM—both in terms of its existence and directional characteristics—has yet to be thoroughly investigated. 
Therefore, to verify and better understand the gradient WM functional variability, we performed a more detailed regional analysis.

\subsection{Regional WM Variability Gradient}
\label{sec:roi_res}


\begin{figure}[htbp]
    \centering
    \centerline{\includegraphics[width=0.8\linewidth]{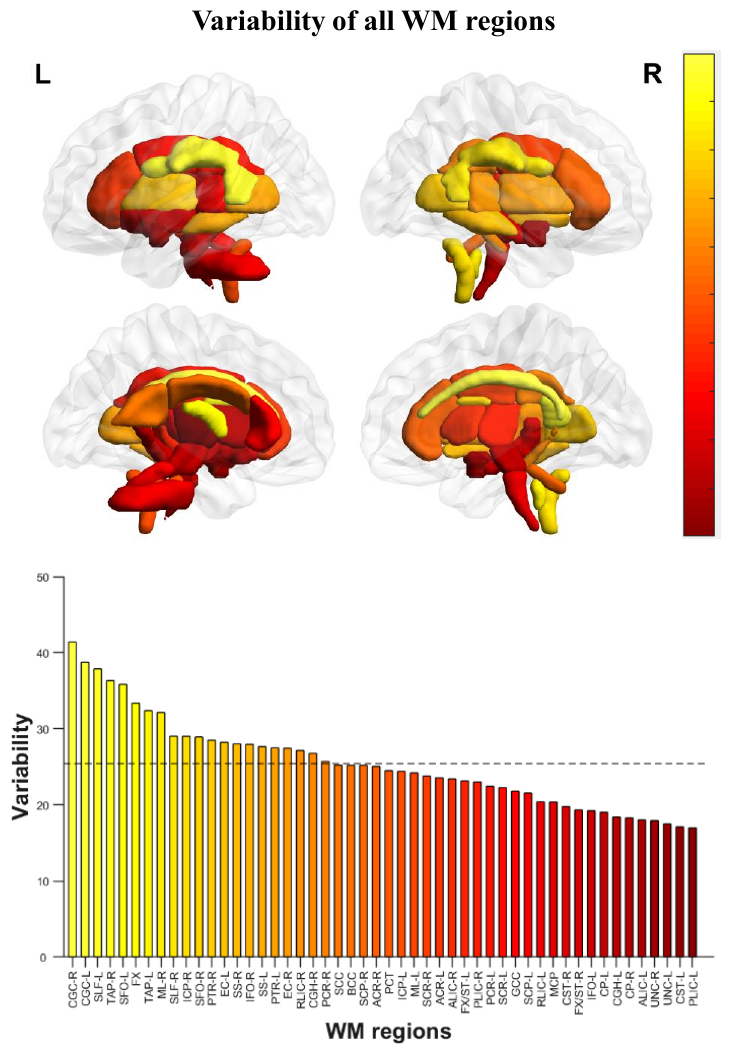}}
    \caption{\textbf{Functional variability of all WM regions in ICBM-DIT-81 atlas.}
    After computing fALFF $I_{diff}$s for 50 WM regions, we visualized the results in both lateral and medial brain views and sorted the regions in descending order of fALFF $I_{diff}$ using a bar chart. In terms of spatial organization, WM regions with high functional variability are predominantly found in more superficial areas, whereas regions with lower variability are located deeper within the WM.}
    \label{fig:idiff_region}
\end{figure}

\begin{figure}[htbp]
    \centering
    \centerline{\includegraphics[width=0.8\linewidth]{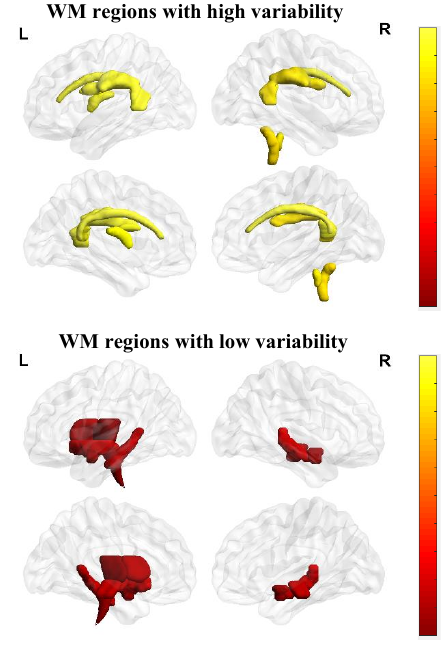}}
    \caption{\textbf{ WM regions with high and low functional variability.} The top 10 and bottom 10 WM regions in functional variability were identified and visualized in lateral and medial views. Clear distinctions in anatomical structure and functional specialization are observed between these two categories of WM regions. Spatially, regions with higher variability are typically positioned closer to the cortex, while those with lower variability are located near the brainstem.}
    \label{fig:idiff_rhigh_low}
\end{figure}



To understand the pattern of functional variability in WM regions, we replaced the group-level mask in Fig.~\ref{fig:schematic} (a) with regions in ICBM-DTI-81 WM atlas and calculated the fALFF $I_{diff}$ for each region. Subsequently, we examined the functional variability and spatial distribution of these regions, as illustrated in Fig.~\ref{fig:idiff_region}.
Relatively shallow WM regions, such as the cingulum (CGC), display much higher functional variability compared to deep periventricular regions like the corticospinal tract (CST). 
This result provides preliminary evidence for a spatial gradient of functional variability across WM.


\begin{table}[htb]
    \centering
    \caption{\textbf{Categories and quantities of WM regions with high and low variability.}}
    \begin{tabular}{l|c|c}
        \hline
        Region category & High variability regions & Low variability regions \\
        \hline
        Association fibers & 6 & 5 \\
        Projection fibers & 0 & {\bfseries 4} \\
        Commissural fibers & {\bfseries 2} & 0 \\
        Tracts in the brainstem & 2 & 1 \\
        \hline
    \end{tabular}
    \label{tab:idiff}
\end{table}

To investigate the spatial variability and its association with WM structure and function, as is shown in Fig.~\ref{fig:idiff_rhigh_low}, we isolated WM regions with high and low functional variability for further analysis.
The ICBM-DTI-81 WM atlas is an atlas based on diffusion tensor imaging (DTI), in which WM is manually segmented into 50 anatomic regions based on fiber orientation information~\cite{atlas2}.
According to directionality and connectivity of the fibers, 50 WM regions are generally divided into four main categories: association, projection, commissural fibers and tracts in the brainstem.
Tab.~\ref{tab:idiff} presents the categorical distribution of WM regions identified as having high or low functional variability.
In Fig.~\ref{fig:idiff_region}, association fiber region cingulum (i. e., CGC-R and CGC-L) demonstrate the most significant functional variability in all WM region, mainly because it is connected to frontal, medial temporal lobes and other cortex region highly correlated with emotions and cognition~\cite{bubb2018cingulum}.
Commissural fibers mainly connect homotopic regions of cerebral hemispheres, while projection fibers are generally referred to bidirectional long fibers between cortex, thalamus, brainstem and spinal cord~\cite{wm_review2}.
The distinct structural connectivities of commissural and projection fiber regions lead to their divergent distributions in high and low variability regions.
Therefore, the WM regional functional variability is highly correlated with structural connectivity, conforming to the coupling relationship between structure and function.

\begin{figure*}[htb]
    \centering
    \centerline{\includegraphics[width=0.8\linewidth]{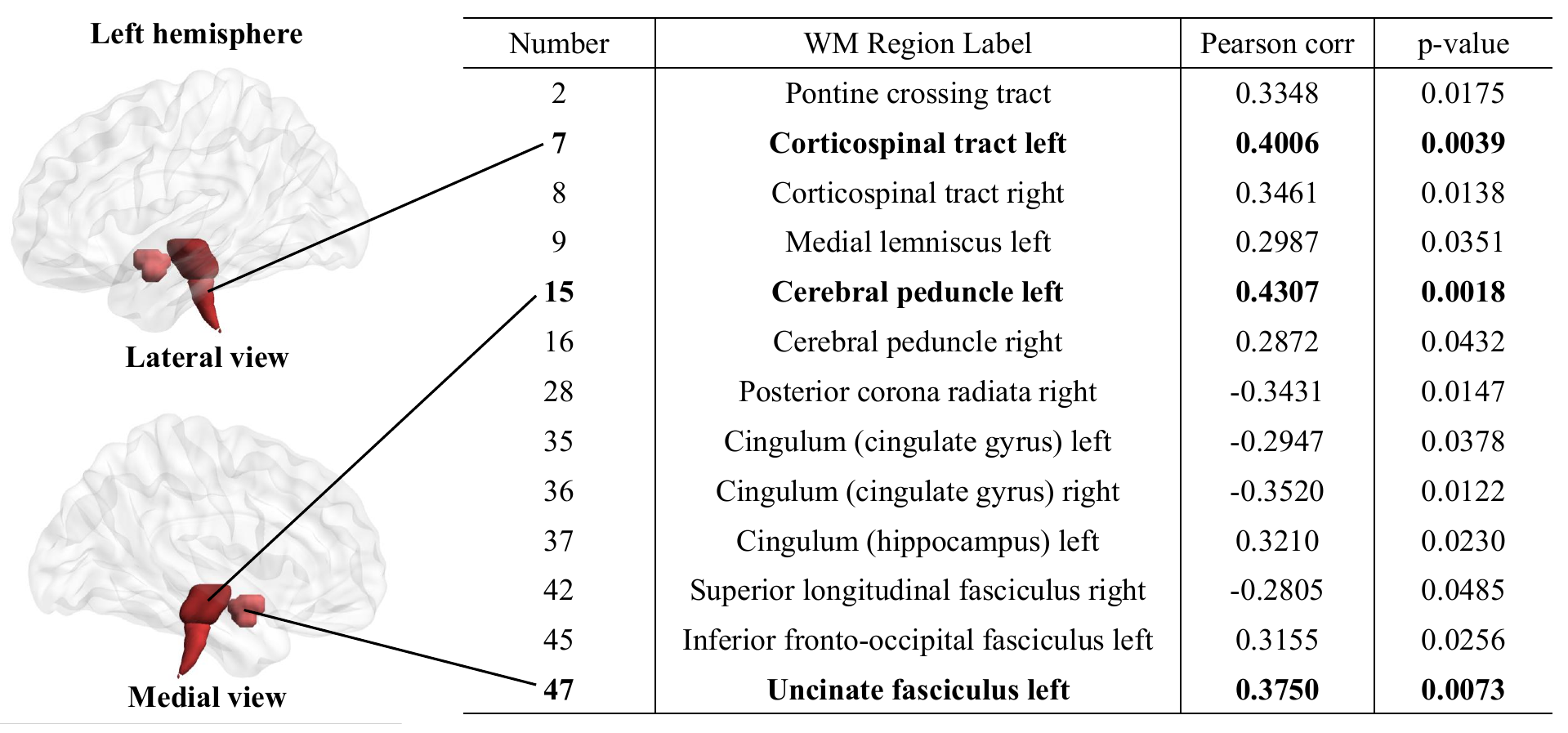 }}
    \caption{\textbf{Pearson Correlations between WM $I_{diff}$ and relative distance across different reference regions (p $<$ 0.05), along with lateral and medial views of 3 notable reference regions.} We computed the relative distances using each of the 50 WM regions as the reference and analyzed the correlations between the relative distances and $I_{diff}$ values. 
    Relative distances to reference regions with p $<$ 0.05 significance are considered statistically correlated with $I_{diff}$ values, with corresponding regions listed in the table. In the 3 reference regions that reach the significance level of p $<$ 0.01, 2 regions coincide with the brainstem, while the third one is nearby. This finding confirms that the functional variability shows the most significant gradient from the brainstem to other WM regions, aligning with our initial hypothesis.}
    \label{fig:falff_idiff_gradient}
\end{figure*}

\begin{figure*}[htb]
    \centering
    \centerline{\includegraphics[width=0.8\linewidth]{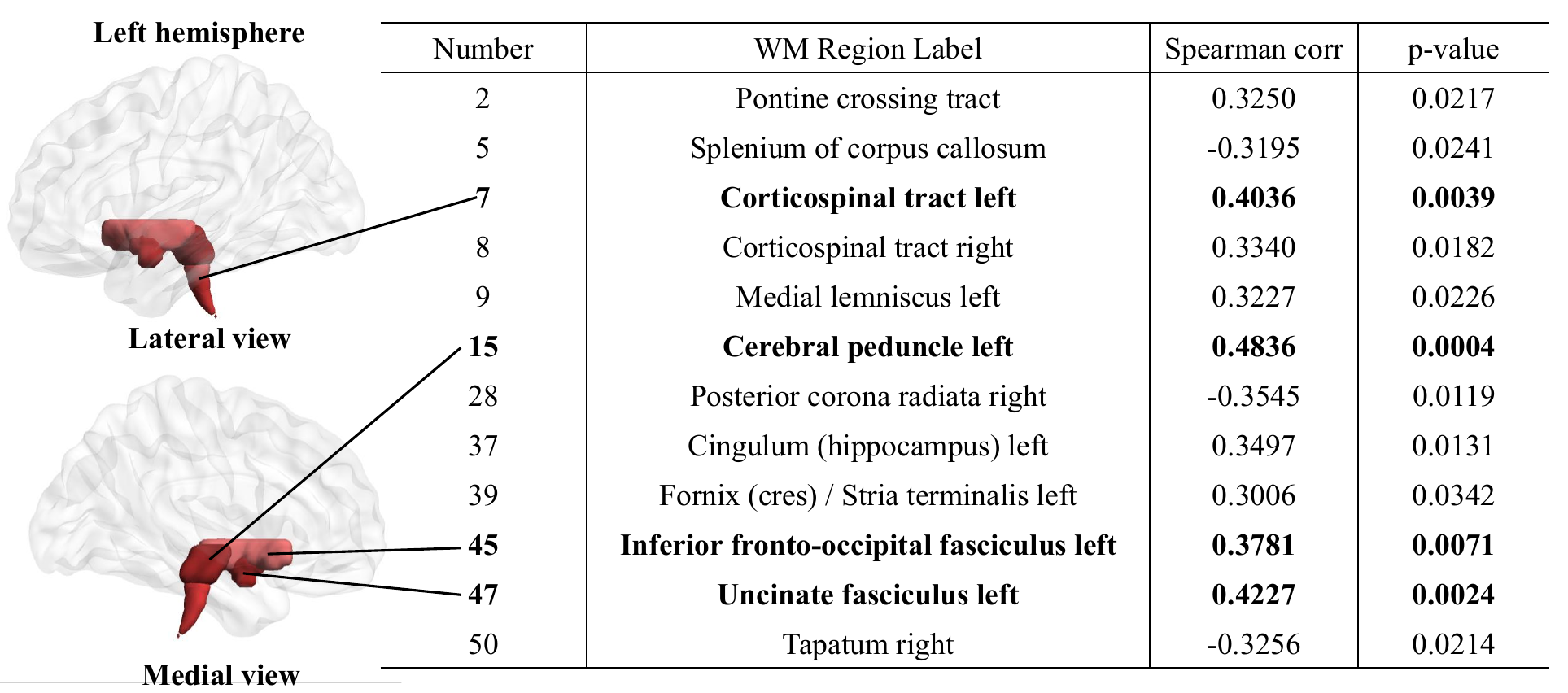}}
    \caption{\textbf{Spearman correlations between WM $I_{diff}$ and relative distance across different reference regions (p $<$ 0.05), along with lateral and medial views of 4 notable reference regions.} We repeated the experiment using Spearman correlation in place of Pearson correlation to assess the robustness of the results. Among the 4 significant reference regions identified using Spearman correlation(p $<$ 0.01), 3 regions overlap with those from the Pearson-based analysis, and the fourth region is also located in close anatomical proximity. The result reinforces our hypothesis of a spatial gradient in WM functional variability.}
    \label{fig:spearman_falff_idiff_gradient}
\end{figure*}

As indicated in Fig.~\ref{fig:idiff_rhigh_low}, WM regions with high functional variability are spatially closer to the cortex, while regions with low functional variability are nearer to the brainstem and thalamus. 
Thus, we hypothesized that from the brainstem to the cortex, fALFF $I_{diff}$ values of WM regions integrally present a spatial gradient distribution that regions closer to the cortex have higher functional variabilities generally, corresponding to brain evolutionary expansion~\cite{hill2010similar}. 
To validate our hypothesis, we analyzed functional variability gradients in various directions. 
Specifically, we employed each region as the reference region and computed the relative distances to all other WM regions. Subsequently, we assessed the Pearson correlation between the relative distances and fALFF $I_{diff}$ values, selecting outcomes with a significance level of p $<$ 0.05, as presented in Fig.~\ref{fig:falff_idiff_gradient}. 
To identify the direction with the most remarkable gradient features, we also visualized 3 reference regions with a significance level of p $<$ 0.01 in Fig.~\ref{fig:falff_idiff_gradient}. 
The experimental results proved the existence of a prominent functional variability gradient from the 3 reference regions to other regions. 
Except for a neighboring region, 2 regions overlap with the brainstem, thereby validating our hypothesis of gradient initially.
Since Pearson correlation primarily captures linear relationships and is sensitive to outliers, we additionally employed Spearman correlation, which emphasizes monotonic associations, to validate our hypothesis.
The experiment was repeated using Spearman correlation in place of Pearson correlation and the results are shown in Fig.~\ref{fig:spearman_falff_idiff_gradient}. Except for 3 significant reference regions consistent with previous results, the fourth was spatially adjacent to them.
Moreover, the significant increase in correlation coefficients indicated a consistent  monotonic association between relative distance and functional variability.
Consequently, consistent findings across both Pearson and Spearman correlation jointly validated our hypothesis that WM functional variability demonstrates a spatial gradient ascending from the brainstem to the cortex. 

\section{Conclusion}
\label{sec:conclusion}
In this paper, we propose a novel method to quantify WM functional variability by estimating differential identifiability based on fALFF similarity. By comparing the differential identifiability of fALFF and FC, we find that fALFF is superior in capturing functional variability. 
Our evaluation of the functional variabilities in both WM and GM reveals a similar overall pattern, though WM exhibits slightly lower variability than GM.  
Subsequent regional analysis uncovers a relationship between WM functional variability and structural connectivity, with commissural fiber regions showing greater variability than projection fiber regions. 
Finally, our most significant contribution is the discovery that WM functional variability demonstrates a spatial gradient ascending from the brainstem to the cortex, consistent with patterns of evolutionary expansion. 
These findings provide novel evidence for the significance of WM function and  support the methodological validity of our method for assessing functional variability across brain regions.

Despite its strengths, our work has some limitations. The current analysis of regional functional variability was confined to WM, but it can be feasibly extended to whole-brain investigations. 
Furthermore, although we identified a spatial gradient of variability in healthy individuals, its alteration in pathological conditions remains an open question for future research. 
In future work, we aim to address the aforementioned limitations. Specifically, we will incorporate GM functional variability to investigate the gradient of variability across both WM and GM. Additionally, we plan to incorporate data from WM-related neurological disorders to explore potential clinical biomarkers within the gradient of variability.

\section{Acknowledgement}
This work was supported by National Natural Science Foundation of China (grant No. 62376068, U24A20340, 32361143787, 82302167), by Guangdong Basic and Applied Basic Research Foundation (grant No. 2023B1515120065), by Guangdong S\&T programme (grant No. 2023A0505050109), by Shenzhen Science and Technology Innovation Program (grant No. JCYJ20220818102414031).

\bibliographystyle{IEEEtran}
\bibliography{refs}

\end{document}